\documentclass[12pt]{article}
\usepackage{graphicx}
\usepackage{amssymb}
\usepackage{subfigure}
\begin{document}

\title{Polygons on a Rotating Fluid Surface}
\author{Thomas R. N. Jansson, Martin P. Haspang, K{\aa}re H. Jensen, \\
Pascal Hersen and Tomas Bohr\\
\small Physics Department, The Technical University of Denmark, 2800 Kgs. Lyngby \\
\small and \\
\small The Niels Bohr Institute, Blegdamsvej 17, 2100 Copenhagen {\O}}
\maketitle

\begin{abstract}
We report a novel and spectacular instability of a fluid surface in a rotating system. In a flow driven by rotating the bottom plate of a partially filled, stationary cylindrical container, the shape of the free surface can spontaneously break the axial symmetry and assume the form of a polygon rotating rigidly with a speed different from that of the plate. With water we have observed polygons with up to 6 corners.  It has been known for many years that such flows are prone to symmetry breaking, but apparently the polygonal surface shapes have never been observed. The creation of rotating internal waves in a similar setup was observed for much lower rotation rates, where the free surface remains essentially flat \cite{Lopez2002}-\cite{Lopez2004}. We speculate that the instability is caused by the strong azimuthal shear due to the stationary walls and that it is triggered by minute wobbling of the rotating plate. The slight asymmetry induces a tendency for mode-locking between the plate and the polygon, where the polygon rotates by one corner for each complete rotation of the plate. 
\end{abstract}

\begin{figure}
\begin{center}
\includegraphics[width=0.33\hsize]{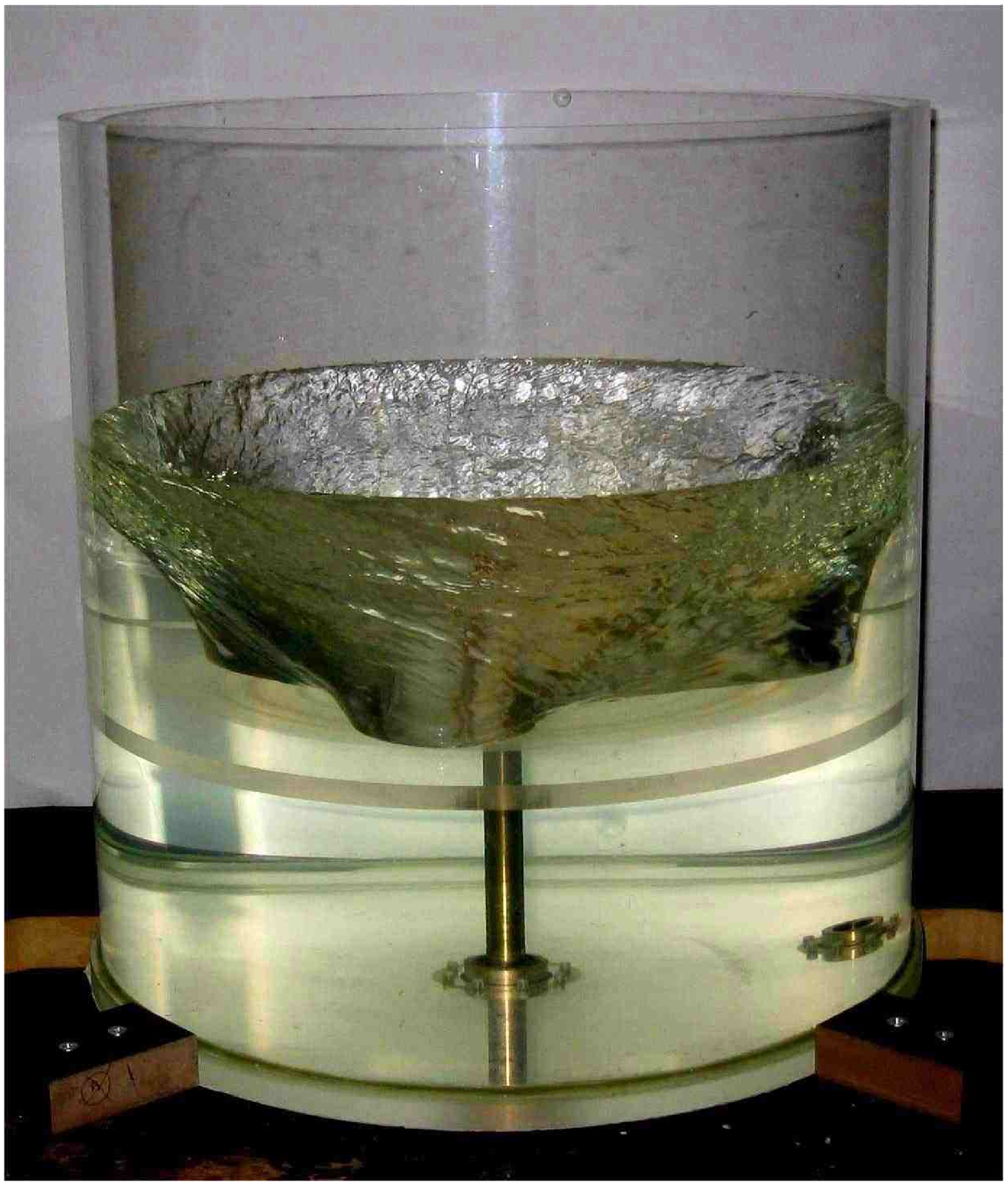}
\includegraphics[width=0.65\hsize]{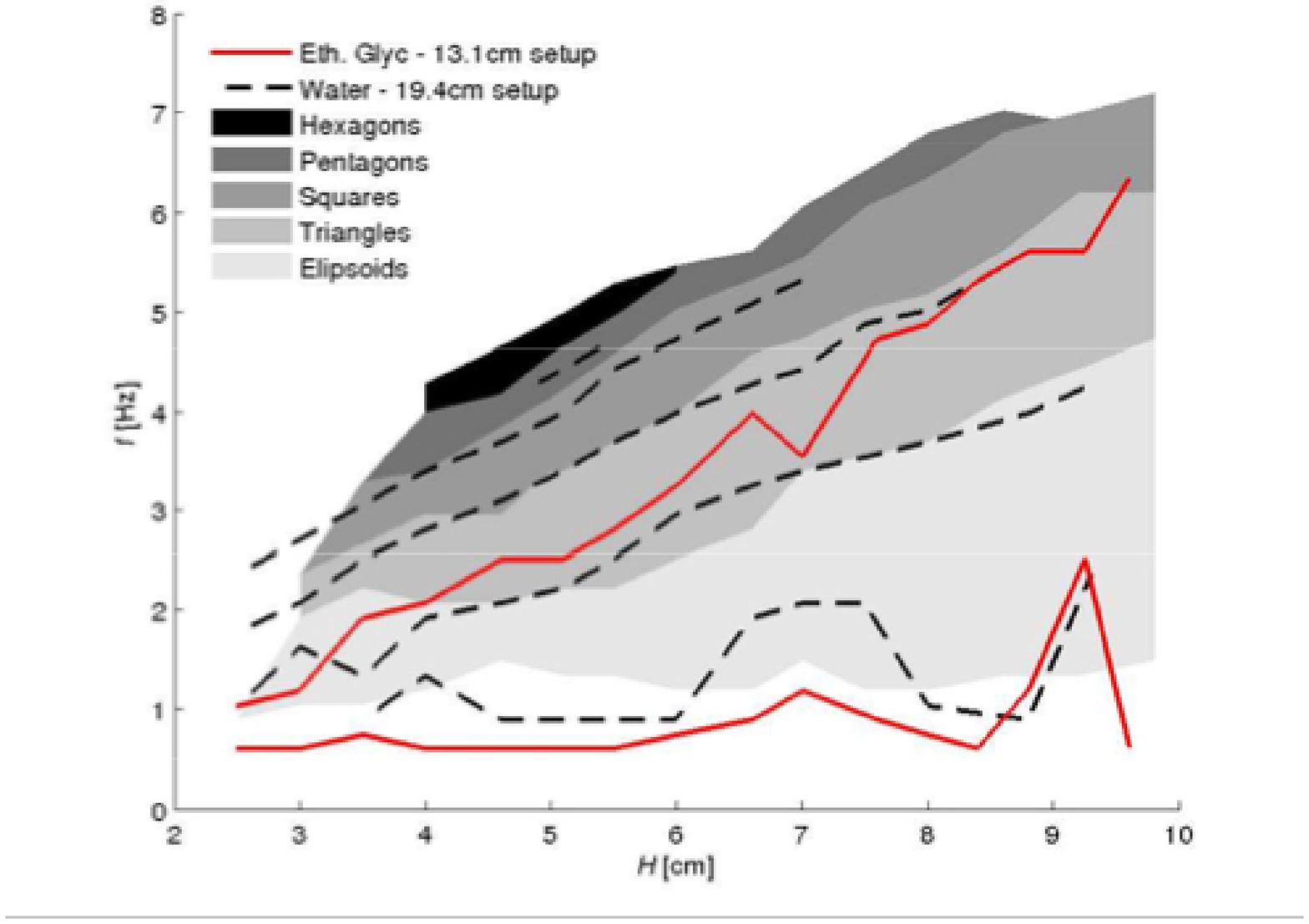}
\includegraphics[width=\hsize]{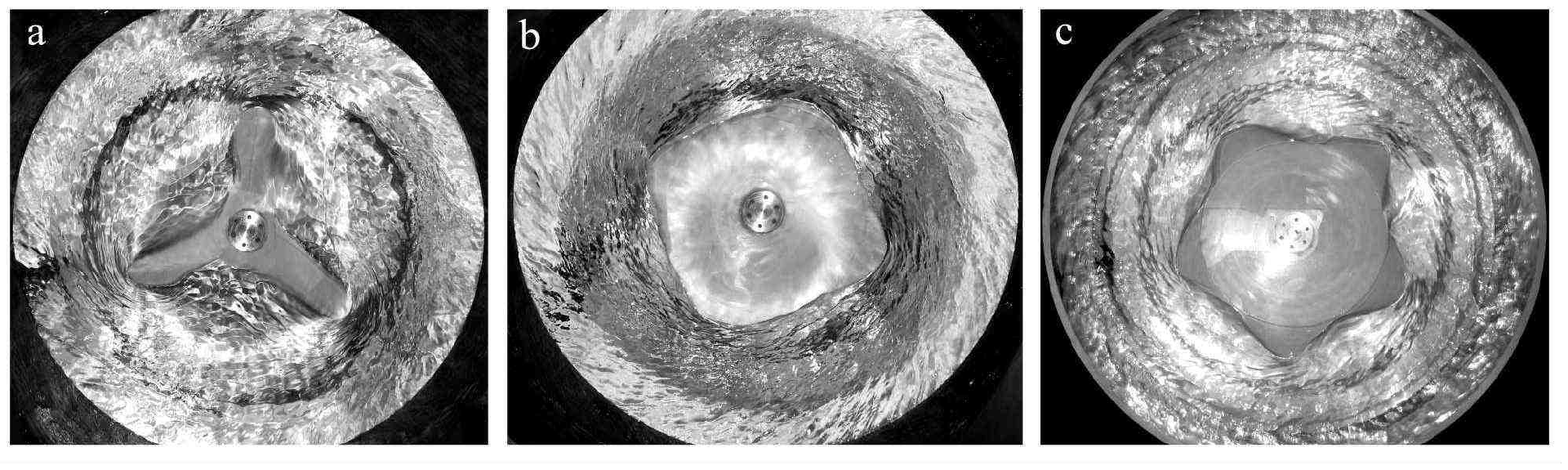}
\caption{
\textbf{Top (left):} Setup consisting of a stationary plexiglass cylinder of radius 19.4 cm with a circular plate that is rotated by a motor. Water or ethylene glycol is filled to the level $H$ above the plate. At sufficiently large rotation frequencies $f$ the axially symmetric surface becomes unstable and assumes the shape of a regular, rigidly rotating polygon, in this case a triangle. \textbf{Top (right):} Phase diagram for  ``polygons" on the surface of a fluid on a rotating plate. The different gray scales correspond to different polygon states as explained in the legend, observed in the smaller setup (with radius 13.1 cm). The dashed lines are similarly  transition lines for transitions between various polygon states, observed for the larger setup (with radius 19.4 cm). The first line (from the bottom) marks the transition
$0\to2$ corners, the second is the  $2\to3$  transition etc. The bottom (top) red
line is the transitions $0\to2$ $(2\to3)$ in the smaller setup using ethylene glycol. We do not see polygons with $N>3$ in ethylene glycol. For most polygons with $N > 2$, the center is dry (as on pictures below). For most of the ellipsoids ($N=2$) and some of the triangles ($N=3$) the surface deformation is milder and there is no dry center. In the white region of the diagram (or below the lowest dashed and red line) the states are circular.  In the upper white region (large $f$) the shapes become very noisy and rigidly rotating polygons cannot be resolved. The phase diagram has been obtained by slowly increasing the frequency at a given height. When the frequency is reversed, some hysteresis is observed, which moves the lines on the order of 0.25 Hz downward.
\textbf{Bottom:} Typical examples of polygons in water, as seen from above: ``Triangle" (a), ``Square"  (b), ``Pentagon" (c). For more pictures and a video of the transition go to www.physics.dtu.dk/$\sim$tbohr/RotatingPolygon.}
\label{fig.block2}
\end{center}
\end{figure}

The experiment  consists of a stationary cylindrical container in which a circular plate is rotated by a motor. Both cylinder and plate are made of plexiglass. Water is filled to the level $H$ above the rotating plate (Fig.~\ref{fig.block2}(top, left)).
When the plate is set into rotation the centrifugal force presses the fluid outward, deforming the free surface. When the rotation rate $\Omega$ becomes sufficiently large, the axial symmetry of the free surface is spontaneously broken and large, time dependent deformations appear. This can result in stable, rigidly rotating surface shapes in the form of regular polygons. Typical examples of polygons in water with $N$ between 3 and 5 are shown below in Fig.~\ref{fig.block2}(bottom).
The surprising and dramatic nature of the transition to the polygon states is best appreciated by looking at the time evolution provided in the video at www.physics.dtu.dk/$\sim$tbohr/RotatingPolygon.

For a given fluid and cylinder, the two control parameters defining the state of the system are the rotation frequency $f = \Omega/2 \pi$ of the plate and the height $H$ of the undisturbed fluid layer. The phase diagram (Fig.~\ref{fig.block2}(top,right)) shows the state i.e. the number of corners $N$ in the polygon state as function of the two control parameters. 
We have used two different setups. In the smaller one the radius of the cylinder is $R=13.1$ cm and the rotating plate is driven by a vertical shaft coming from above. The states for this setup are marked by gray scales. In the larger setup  the radius is $R=19.4$ cm and the shaft comes from below. Here the borders between various polygon states are marked by dashed lines. Aside from water, the experiments have been carried out with ethylene glycol with a viscosity of around 15 times larger than water and the red lines mark transitions between polygon states for ethylene glycol. It is seen that the polygon states basically fill out a whole region of the phase diagram where the transition lines take the system directly from one kind of polygon to another \cite{foot}. 

The phase diagram (Fig.~\ref{fig.block2}(top,right)) is surprisingly simple: the higher the rotation frequency (at fixed $H$) the more corners, and the larger the height (at fixed $f$) the fewer corners. In fact the transition lines $f_N(H)$ between various polygon states are roughly straight lines, i.e. of the form $f_N \sim \alpha_N H + \beta_N$. A surprising feature is the close correspondence between the two setups, although they differ markedly in radius (roughly by a factor 1.5). With ethylene glycol we only observe polygons with $N \leq 3$, but for these polygons the effect of viscosity on the transition lines is surprisingly small. We also tested the dependence on surface tension by injecting a surfactant (detergent) in to the flow and the variations were very slight.

\begin{figure}
\begin{center}
\includegraphics[width=0.30\hsize]{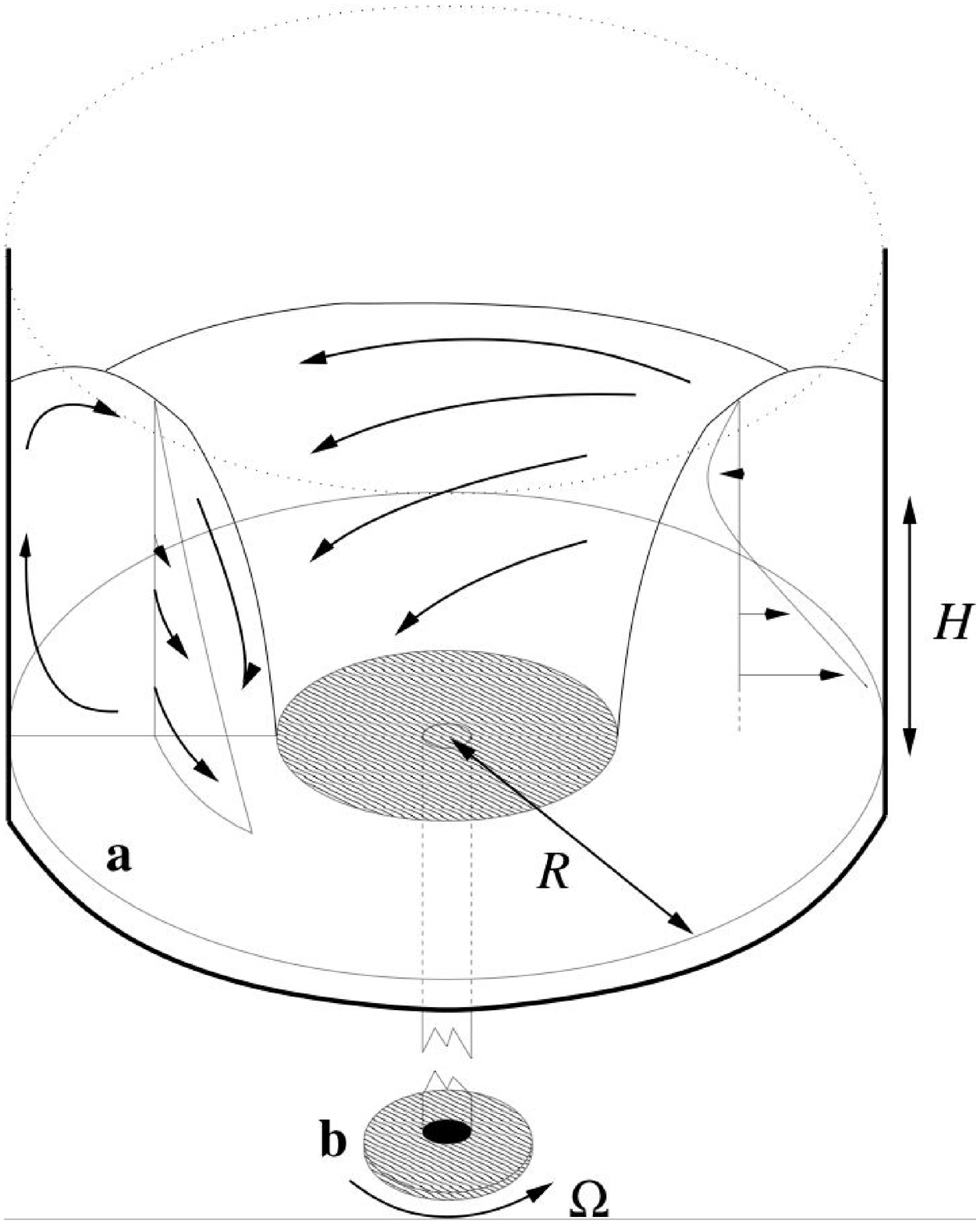}
\includegraphics[width=0.225\hsize]{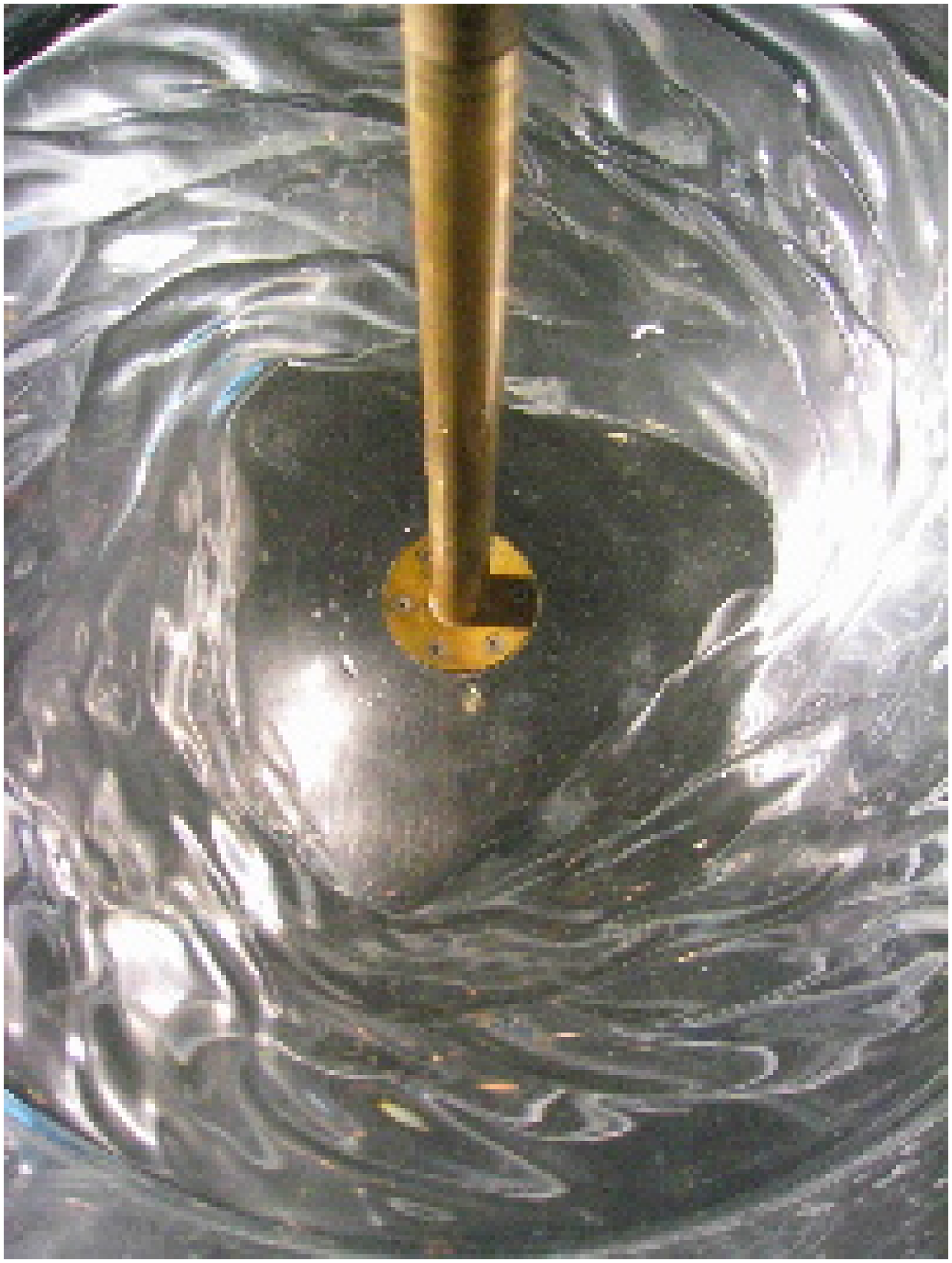}
\includegraphics[width=0.4\hsize]{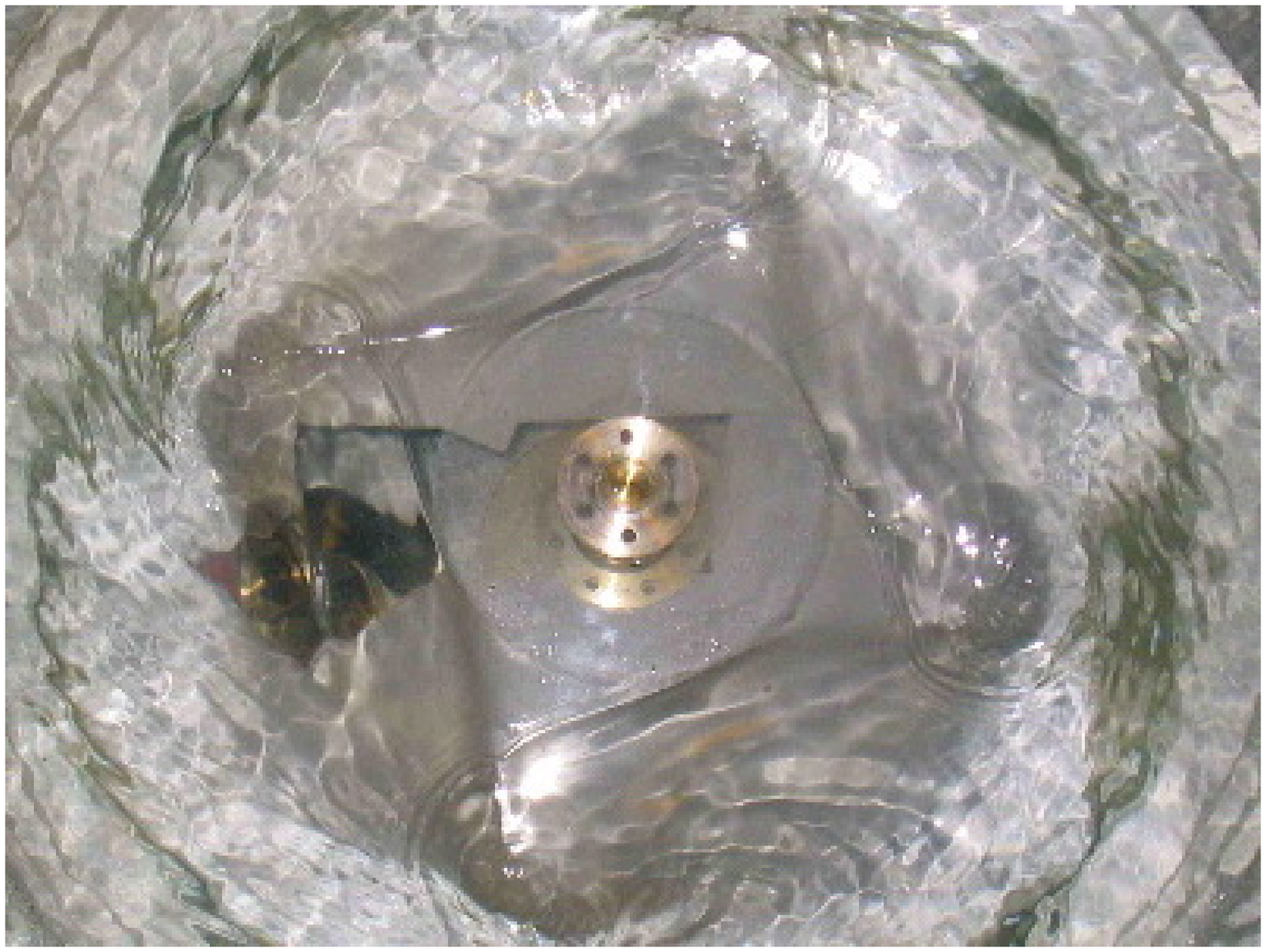}
\caption{{\bf Left:} Sketch of the flow in the axially symmetric basis state on top of which the rotating polygons form. The plate (a) is rotating with angular velocity $\Omega$, driven by a motor (b), the fluid above the plate is pressed radially outward and a secondary flow developes as sketched. {\bf Center:} Triangle formed spontaneously on the surface of ethylene glycol in a cylindrical container over a rotating plate (small setup). Note the ``G\"ortler vortices" spiralling on the surface. {\bf Right:} A rotating square on water. Note the vertical vortices just outside of the corners.}
\label{vortices}
\end{center}
\end{figure}

A triangle state on an ethylene glycol surface is shown in Fig.~\ref{vortices}(center) and we clearly see a pattern of spiraling vortices on top of the polygon structure. These vortices are indicative of the secondary flow shown in Fig.~\ref{vortices}(left) and we interpret these vortices as G{\"o}rtler vortices along the curved streamlines of the flow  \cite{Lugt}. Their width should be determined by the viscous boundary layer, which is proportional to $\sqrt{\nu}$ and thus around 4 times larger in ethylene glycol than in water.

The rotation rate of the polygons is considerably lower than that of the plate and varies with the latter in a complicated way.  On the background of a slow, power-like variation there is a tendency for mode-locking such that the polygons rotate by one corner for every complete rotation of the plate. In the case of ellipsoids (``bi-gons") we have even observed mode-locking in the ratio 2/3.
This must be related to the slight wobbling of the plate, which, although minute, breaks the axial symmetry \cite{foot2}. 

The flow structure in the axially symmetric basis state is sketched in Fig~\ref{vortices}(left). The flow over an infinite rotating plate (the von Karman flow) has an outward radial component. The fixed cylinder wall modifies this flow and introduces a region of strong shear in the azimuthal velocity as well as upwelling near the walls with a subsequent reinjection inward along the free surface \cite{Lugt}-\cite{Mory}. Qualitative properties of such flows have been analysed in \cite{Brons} and they are very important in the context of vortex break down \cite{Spohn} and the general spin-up problem \cite{duck}. The details of the flow are very complicated due to the singular corners, where the rotating plate meets the fixed wall, and which play an important part in the generation of the secondary flow \cite{Ungarish}.  The Reynolds number Re=$\Omega R^2 /\nu$ is in the range between $10^5$ and $1.5 \times 10^6$ for our experiments in water and these large values mean that the flow is actually turbulent and hard to visualize. In earlier work, \cite{Lopez2002}-\cite{Lopez2004}, the symmetry breaking of similar flows has been studied at much lower Reynolds numbers (a few thousand), where the free surface remains essentially undeformed. In \cite{Lopez2004} detailed numerical simulations were performed for this case (i.e. neglecting the variation of the free surface shape) and a transition to a state with a rotating wave with $N=3$ was observed. In this work, the dependence on viscosity was strong and the Reynolds number was a relevant control parameter. 

We believe that our polygon states are interesting new members of a fascinating class of systems, where spontaneous breaking of the axial symmetry leads to simple stationary or rigidly rotating shapes. Earlier examples are hydraulic jumps in an axially symmetric setting \cite{Ellegaard} and instabilities of a shear flow in a thin layer of fluid with differential rotation \cite{Konijnenberg}. In the latter experiment a circular shear layer is generated in a thin rotating fluid layer by letting the inner part of the container rotate at a different rate. Visualization with dye shows strings of vortices along the edge of the inner part and the shapes of the inner boundary of the dye has the form of polygons. In our experiment a shear layer indeed exists due to the no-slip condition on the stationary cylinder wall and could indeed lead to instability of the classical 
Kelvin-Helmholtz/Rayleigh type \cite{Heijst,Konijnenberg}.  In some cases we actually observe vortices close to the corners of the polygons. This is shown very clearly in Fig.~\ref{vortices}(right), where vortices are seen outside each of the four corners. We therefore believe that vortex formation and interaction is very important for the development and stability of the final state.

\begin{thebibliography}{99}
\bibitem{foot}
Close to the transition lines, in particular $2 \rightarrow 3$, transition states do exist, which are not rigidly rotating polygons, but can e.g.  periodically change from 2 to 3 corners. For most ellipsoid and some triangles, the surface deformations are so mild that there is no dry center on the rotating plate. The transition states are usually related to the appearance of a dry center in one of the states.
\bibitem{foot2}  The angle of deflection of the plate from horizontal is around 0.9$^\circ$ in the small and 0.15$^\circ$ for the large setup.
\end{thebibliography}
\end{document}